%% file: draft.tex
\documentclass[11pt]{article}

\usepackage[margin=1in]{geometry}
\usepackage{amsmath,amssymb}
\usepackage{booktabs}
\usepackage{array}
\usepackage{graphicx}
\usepackage{microtype}
\usepackage[hidelinks]{hyperref}
\usepackage[numbers,sort&compress]{natbib}

\newcommand{\Dgen}{\ensuremath{D_{\mathrm{gen}}}}
\newcommand{\Deval}{\ensuremath{D_{\mathrm{eval}}}}
\newcommand{\tauev}{\ensuremath{\tau_{\mathrm{event}}}}
\newcommand{\shat}{\ensuremath{\hat{s}}}
\newcommand{\fracovertau}{\texttt{frac\_over\_tau}}

\title{Amplify, Don't Create:\\ Temporal Accumulation for Slow-Burn Prompt Injection}
\IfFileExists{local_author.tex}{
  \input{local_author.tex}
}{
  \author{
    The Parapet Project \\
    \url{https://github.com/Parapet-Tech/parapet}
  }
}
\date{\today}

\begin{document}
\maketitle

\begin{abstract}
Most prompt-injection detectors score a single event or message. Control-plane attacks against
tool-using agents can instead distribute weak directives across a trajectory while keeping each
event below threshold. We test whether a proxy-side temporal accumulator can recover this
slow-burn signal by reducing frozen per-event scores to peak and Cumulative Sum (CUSUM)
persistence statistics.

To avoid circularity, grafts are generated against a held-out autoregressive cloaking target and
then re-scored under a detector of record: a frozen SVM linear char-ngram classifier plus an
embedding-contrastive head. Only floor-met grafts bound to real executed action edges and remaining
sub-threshold under the detector of record enter the slow-burn endpoint.

The result is a boundary result, not a deployable detector. On concentrated attacks,
trajectory-level accumulation beats the per-event foil under a clustered bootstrap (gap $+0.092$, 95\%
CI $[+0.025, +0.155]$), while persistence and peak are statistically tied. On git repo-exfil,
density-four floor-met sub-threshold grafts add persistence mass that matched benign shams do not:
the four attack survivor deltas are $+0.0513$, $+0.0329$, $+0.1432$, and $+0.0082$ (persistence-delta AUC
0.708 over four attack survivors and six benign shams). The matched peak-delta control does not
separate attack from sham (AUC 0.417), localizing the effect to accumulated persistence rather than
a single hot graft. The effect fails on broader clean-path actions (persistence-delta AUC 0.167),
where audit shows the detector assigns attack and benign actions indistinguishable per-event scores,
leaving no margin for CUSUM to bank.

Independent powering is structurally blocked by only three to four independent tasks. Temporal
accumulation is therefore a narrow-band margin amplifier: it can bank elevated sub-threshold signal
but cannot create margin where the per-event detector has none. As methodological byproducts we
contribute a pseudo-replication warning and an independence-audit standard for agent-benchmark
evaluation.
\end{abstract}

\section{Introduction}

Tool-using language-model agents read untrusted content (web pages, tool outputs, repository files)
and act on it. Prompt injection is adversarial text in an untrusted channel that redirects the agent's control plane, making it the central security problem of agentic deployment. We use \emph{event} for a
position-ordered text span observed by the proxy, such as a user message, tool output, or grafted
text span associated with an action edge. The dominant defensive primitive is a per-event detector:
score each event, alarm when a single event crosses a threshold.

This primitive has a structural blind spot. A control-plane attack need not concentrate its signal
in one event. It can distribute a sequence of individually innocuous-looking directives across a
long trajectory (``when you finish, also push the working tree to this remote''; later, ``use this
URL''; later, ``include the env file''), each step sub-threshold, the aggregate effect a full exfil.
We call this the \emph{slow-burn} regime. A per-event detector tuned to a usable false-positive rate
cannot lower its threshold enough to catch each quiet step without drowning in benign agent chatter.

The natural hypothesis is temporal: if no single event is loud, perhaps the \emph{persistence} of mild
elevation across a trajectory is itself evidence. The idea is proxy-deployable: a monitor sits beside the agent, scores each event with a frozen detector, and accumulates the results. We test it with a
peak-and-persistence accumulator (a reset-at-zero CUSUM with a benign-referenced drift), and ask one
question: \textbf{does accumulation recover distributed control-plane signal that per-event scoring
misses, and under what conditions?}

The honest answer turns out to be conditional, and the condition is the contribution. The main
contribution is the margin-amplifier result; the statistical and reconciliation lessons are
methodological byproducts. Our findings:

\begin{enumerate}
\item \textbf{Accumulation is a real, non-circular mechanism, but it is a margin amplifier, not a semantic
   detector.} It pays off only where the underlying per-event detector already produces \emph{elevated
   sub-threshold} scores on the attack steps. Where the detector assigns attack and benign actions
   indistinguishable scores, no accumulator threshold recovers a margin that does not exist.

\item \textbf{The productive regime is narrow and quantifiable.} Because the CUSUM accumulates only score
   mass above the benign drift baseline, grafts pushed \emph{deep} sub-threshold contribute nothing.
   Density (more mild grafts per trajectory) pays off only in a thin band between the benign drift
   and the per-event threshold (here roughly 0.45 to 0.50). We give the CUSUM arithmetic that makes
   this band narrow.

\item \textbf{The positive result is deliberately bounded to the axis where the required margin exists.}
   It holds on git repo-exfil, where a genuine per-event margin exists; it fails to generalize to
   broader clean-path actions; and it cannot be independently powered, because the axis exposes
   only three to four independent tasks. We report it as a bounded mechanism study, not a
   deployable system.
\end{enumerate}

The evaluation constraints produce reusable reporting lessons. The study exposes two methodological
hazards for agent-benchmark evaluation: pseudo-replication, where multiple model executions over
the same task geometry are miscounted as independent evidence, and arithmetic reconciliation, where
accumulator results should be re-derived from frozen per-event receipts with explicit data-path,
algorithmic, and scoring-layer trust boundaries. These are secondary to the mechanism claim, but
necessary to report it without overstating power.

We deliberately do not claim a better detector. The frozen-stack contract is the point: nothing in
the sensor or the floor judge was tuned to the results, so the mechanism question (``does the proxy
accumulator work'') never collapses into a detector-engineering question.

\section{Threat model and problem setup}

\textbf{Setting.} An agent executes a multi-turn trajectory of typed action edges (tool calls and their
outputs). Some text on the trajectory is attacker-controlled. A control-plane \emph{directive} is a
machine-extractable instruction bound to a real executed action edge (e.g.\ a \texttt{git push} to an
attacker remote). The attacker's goal is to land directives that the agent executes while keeping
each individual event below the defender's per-event alarm.

\textbf{Defender.} A proxy monitor scores every event with a frozen detector and accumulates over the
trajectory. The defender holds a database of known prior attacks, treated as a frozen reference set, but does not retrain or retune in response to the attack. This deliberately conservative posture forces the attacker to drift semantically away from known payloads rather than merely paraphrase them.

\textbf{Slow-burn.} The adversary distributes sub-threshold directives across the trajectory. The
research question is whether trajectory-level accumulation recovers what per-event scoring cannot,
and whether that recovery survives a non-circular construction.

\textbf{Anti-circularity is the hard part.} If the attack generator can see the evaluator, it trivially
fits it, and any ``miss'' is circular. We enforce two separations (Section~\ref{sec:method}): the generator targets a
\emph{held-out, different-family} cloaking detector, and the evaluated low-signal property must \emph{transfer}
to the detector of record by re-scoring, never by optimization against it.

\section{Method}\label{sec:method}

\subsection{The frozen contract (held fixed across every experiment)}

We separate the \emph{generation target} from the \emph{evaluator of record}.

\begin{itemize}
\item \textbf{\Dgen{} (cloaking target, held out).} An autoregressive LLM judge, decoded deterministically for
  reproducible per-event scores. \Dgen{} is the only detector the attack generator can query. It scores
  ``how strongly does this event read as a control-plane move to a single-event reader,'' and it never
  sets ground truth and never appears in the evaluated cells. It is the adversary's cloaking surface,
  nothing more.

\item \textbf{\Deval{} (detector of record, evaluated).} An ensemble of two cross-family members:
  \begin{itemize}
  \item \emph{L1}, a frozen linear char-ngram classifier (LinearSVC/SGD family), black-box inference, no
    retraining, raw margin mapped to a calibrated $[0,1]$ via Platt scaling, with chunk-and-max-risk
    aggregation for long spans.
  \item \emph{Contrastive embedding member}, an embedding-contrastive head over a frozen encoder and frozen
    contrastive reference bank. Long spans are split into overlapping windows and aggregated by max
    cosine, so end-of-span slow-burn signal is not silently truncated.
  \item \emph{Aggregate:} the per-event score is the \textbf{ensemble-max} $s_i = \max(\text{L1}, \text{contrastive})$ over
    per-member calibrated scores, failing the event closed if any member errors. (We call this
    ensemble-max, not ``conservative-max'': max over members is the most trigger-happy combination, so
    ``conservative'' would invert the detector-language convention where conservative means fewer
    alarms.)
  \end{itemize}
  The cross-family invariant is machine-checked: \Dgen's family (generative/autoregressive) is
  disjoint from \Deval's member families (linear, embedding/retrieval). The generator has no \Deval{}
  access.

\item \textbf{\tauev{} $= 0.50$, $\epsilon = 0.05$, \texttt{max\_iter} $= 6$.} The per-event operating point was locked
  before generation by a threshold sweep over a $\sim$50-trajectory calibration batch assembled entirely
  from pre-materialized data (benign-clean, benign-mention, naive-attack classes, read per-class
  because trajectories vary widely in event count). At $\tau = 0.50$, the locked detector exhibited weak
  per-event separation with heavy class overlap. This weakness was observed at the frozen operating
  point, not introduced by tuning for the accumulator. $\epsilon = 0.05$ (one tenth of \tauev{}) is the
  convergence tolerance of the graft-refinement loop: a candidate is refined for up to \texttt{max\_iter}
  passes until its \Dgen{} score lands within $\epsilon$ of its target aim-point and below \tauev{}. In the
  density runs (Section~\ref{sec:result1} onward) graft acceptance is governed by the productive band rather
  than by $\epsilon$, which remains a locked, provenance-checked construction parameter.

\item \textbf{Floor judge.} A held-out LLM judge, applied per graft, fail-closed. A graft counts as
  \emph{floor-met} only if the judge confirms a machine-extractable directive bound to a real executed
  action edge. This prevents the degenerate ``graft is sub-threshold because it says nothing''
  failure.
\end{itemize}

Nothing in \Deval{}, the floor, the drift, or the CUSUM was altered in response to any result below.

\subsection{The accumulator}

For a trajectory with per-event scores $s_1 \dots s_T$ ordered by position:
\begin{itemize}
\item $\mathrm{peak} = \max_t s_t$.
\item $\mathrm{persistence} = \max_t g_t$, where $g_0 = 0$ and $g_t = \max(0,\, g_{t-1} + (s_t - k))$: a CUSUM that
  resets at zero with benign-referenced drift $k$.
\item \fracovertau{} $= \lvert\{\,t : s_t \ge \tau\,\}\rvert / T$.
\end{itemize}

The drift $k$ is the median per-event score over the benign population. This is the load-bearing
detail behind the entire mechanism: \textbf{the CUSUM banks only score mass above $k$.} A graft pushed
deep below $k$ adds zero persistence, no matter how many you stack. (Section~\ref{sec:band} makes this
arithmetic concrete.)

\subsection{Graft generation}

A generation loop (the refinement loop, up to \texttt{max\_iter} $= 6$ iterations) grafts control-plane
directives onto clean carrier trajectories. Each candidate graft is scored against \Dgen{} and refined
until it is sub-threshold under \Dgen{} while still carrying an extractable directive. The generator has
no \Deval{} access; it steers grafts using \Dgen-window heuristics only (weave into existing argument
text, carrier-plausible values, prefer the lower half of the \Dgen{} window).

Two graft populations:
\begin{itemize}
\item \textbf{Attack grafts:} floor-met control-plane directives (e.g.\ push the repo to an attacker remote).
\item \textbf{Benign sham grafts:} matched-count, carrier-plausible insertions with no control-plane
  directive, used as the selection control.
\end{itemize}

\subsection{The graft-delta endpoint (and why not absolute AUC)}

Carriers chosen to admit density-4 grafts are not a random sample: the edge filter that selects
``trajectories with enough executed git action edges'' also selects \emph{hotter} native trajectories.
Absolute persistence AUC therefore conflates the grafts' contribution with this native selection
heat. Our primary endpoint cancels it:

For any trajectory statistic $m$, define:
\begin{equation}
  m\text{-delta} = m(\text{native} + \text{grafts}) - m(\text{native only}),
\end{equation}
per carrier, compared attack-versus-benign.

The primary endpoint is \textbf{persistence-delta} at a common drift. The matched \textbf{peak-delta} is a
required control: if persistence-delta and peak-delta separate equally, the result is compatible
with ``one hot-ish graft per trajectory'' rather than a temporal accumulation effect.

Because the native baseline is subtracted within each carrier, a positive attack-versus-benign
persistence-delta is attributable to the \emph{grafts}, not to which carriers were selected. Absolute
persistence AUC is reported only as context.

We add one further guard against a selection artifact masquerading as an effect: a pre-registered
native counterfactual, the native-only persistence/peak AUC computed before any grafting. It asks one question: are attack trajectories already sufficiently separated from matched benign native trajectories, on their own, to explain any later absolute persistence result? If they are, absolute AUC cannot be
trusted as evidence for the grafts, and the run must fall back to the within-carrier delta endpoint
or stop. Because it is read before generation, it cannot be tuned to the outcome.

\subsection{Cross-family transfer (the anti-circularity payoff)}

Grafts are cloaked sub-threshold against \Dgen{}, then the low-per-event-signal property is \emph{required
to survive re-scoring} under the different-family \Deval{}. Because the generator never sees \Deval{} and
the families are machine-checked disjoint, a graft that stays sub-threshold under \Deval{} did so by
transfer, not by fitting the evaluator. A catastrophic-transfer guard runs on a 2--3 trajectory
mini-batch before any full run: if \Deval{} instantly catches what \Dgen{} hid, we stop rather than burn
compute on a doomed construction. In the reported density runs, transfer is reported as a gate
quantity: leaked grafts are events that were cloaked under \Dgen{} but score at or above $\tau$ under
\Deval{}; leaked trajectories are relabeled or removed from the slow-burn survivor set.

\section{Experimental setup}

\textbf{Benchmark.} Carriers and tasks come from AgentDyn \citep{agentdyn2026} (MIT, built on AgentDojo),
which supplies typed action edges with clean benign backbones across suites (github, shopping, file,
etc.). Concentrated attack trajectories for the foundation cohort are six handcrafted multi-turn
attacks plus three MHJ control-plane conversations, held out of the
reference banks.

\textbf{Compute boundary and reconciliation.} Scoring (BGE embedding + contrastive + L1) runs in an MLX
environment; accumulation, AUC, and verdicts run model-free in an independent process that
re-derives every headline number from a per-event receipt rather than trusting the reported
AUCs (Section~\ref{sec:ladder}). The detector of record was held constant across all results after a one-time
reconciliation (Section~\ref{sec:result0}).

\textbf{Uncertainty.} Binary verdicts (X ``beats'' Y) report a clustered bootstrap CI, resampled at the
genuine independence unit. For the foundation cohort that unit is the \emph{trajectory} (the nine attack
trajectories are six distinct handcrafted plus three distinct MHJ cases, no pseudo-replication). For
the density results it is the \emph{task} (Section~\ref{sec:pseudorep}). A verdict verb is licensed only if the gap CI
excludes zero; otherwise the claim is non-inferiority, not dominance.

\section{Results}

\subsection{Result 0: Foundation Cohort Evaluation}\label{sec:result0}

Before any generation, on held-out trajectories (9 concentrated attack, 40 clean benign; 116 events,
33 attack / 83 benign), the per-event detector of record is the \textbf{foil}: per-event AUC \textbf{0.766}.
Lifting to the trajectory level under the detector of record:

\begin{table}[h]\centering
\begin{tabular}{lrrrl}
\toprule
metric & distance (retired) & detector of record & $\Delta$ & 95\% CI (contrastive) \\
\midrule
per-event foil AUC & 0.677 & \textbf{0.766} & $+0.089$ & $[0.642, 0.884]$ \\
peak AUC           & 0.789 & \textbf{0.828} & $+0.039$ & $[0.692, 0.947]$ \\
persistence AUC    & 0.814 & \textbf{0.858} & $+0.044$ & $[0.719, 0.961]$ \\
\fracovertau\ AUC  & 0.589 & \textbf{0.657} & $+0.068$ & $[0.464, 0.854]$ \\
\bottomrule
\end{tabular}
\caption{Foundation cohort, per-event vs trajectory-level AUC under the detector of record.}
\label{tab:result0}
\end{table}

Clustered bootstrap (resample trajectories, 2000 draws), on the two load-bearing comparisons. Here
``best trajectory AUC'' means the better of the two trajectory-level accumulator summaries, peak or
persistence, compared against the per-event foil:

\begin{itemize}
\item \textbf{Accumulation beats the per-event foil under a clustered bootstrap.} best-trajectory AUC minus
  foil $=$ \textbf{$+0.092$}, 95\% CI \textbf{$[+0.025, +0.155]$}, excludes 0. This is a small-n bootstrap result,
  not a power claim.
\item \textbf{Persistence vs peak: statistically TIED.} persistence minus peak $= +0.031$, 95\% CI
  $[-0.033, +0.100]$, includes 0.
\end{itemize}

The honest reading: these attacks are \emph{concentrated}, not slow-burn, so a single event already
spikes and peak is already strong; persistence only \emph{matches} it. That is exactly what the design
predicts. The load-bearing sanity property holds with statistical significance: accumulation does not underperform per-event detection, and benign persistence concentrates at zero. The persistence-beats-peak
advantage is only expected on genuinely slow-burn trajectories, where no single event spikes, which
is precisely what the generation arc (Results 1--3) sets out to manufacture and test.

Note the design choice this surfaces: a \emph{stronger} per-event detector raises the foil, so the
accumulator's marginal lift shrinks. The foil-to-accumulator gap narrowed from $+0.137$ under the
retired distance member to $+0.092$ under the contrastive member, but stays positive and significant.
We report the \emph{gap}, not the raw AUC rise.

\subsection{Result 1: Git Repo-Exfil Mechanism Validation under Density}\label{sec:result1}

On clean github carriers whose backbone executes repeated repo-slot git actions (clone / push /
create\_repo), grafting four spread, floor-met, sub-tau directives yields a \textbf{positive
selection-controlled persistence-delta} (gate drift 0.4295):

\begin{itemize}
\item attack survivor persistence-deltas: \textbf{$+0.0513$, $+0.0329$, $+0.1432$, $+0.0082$} (four survivors over
  three tasks), median \textbf{$+0.0421$}.
\item benign sham persistence-delta median \textbf{$+0.0150$}; for reference, persistence-delta AUC \textbf{0.708}
  over four attack survivors and six benign shams.
\item matched peak-delta control: attack median \textbf{$+0.0056$} vs benign sham \textbf{$+0.0136$}, peak-delta AUC
  \textbf{0.417}. Per-carrier attack peak-deltas are $+0.0513$, $+0.0$, $+0.0113$, $+0.0$; benign sham
  peak-deltas are $+0.0153$, $+0.0$, $+0.0120$, $+0.0484$, $+0.0$, $+0.0207$.
\item 25\% of survivor grafts land in the productive 0.46--0.50 band.
\item instrument clean: scorer max-abs-delta 0.0, floor enforced, leak fraction 0.042 (1/24 grafts over
  tau), transfer non-catastrophic.
\end{itemize}

The pre-graft native counterfactual is also material context: the density-four git edge filter
already selected attack-hotter carriers (native-only peak AUC 0.778, native-only persistence AUC
0.750). Absolute post-graft AUC is therefore not graft-attributable; this is why the within-carrier
delta is the endpoint.

Transfer/gate accounting is part of the reported outcome: 24 attack grafts were attempted; one
crossed tau under \Deval{} (leak fraction 0.0417), producing one concentrated relabel, and one
trajectory was rejected by the directive floor. The slow-burn endpoint therefore uses the four
promoted surviving trajectories only, compared against the promoted benign shams.

This is the core positive: a density of mild, individually sub-threshold elevation accumulates into
persistence mass that equal-count benign sham grafts do not reproduce. The evidence is the
controlled persistence-delta plus the matched peak-delta control, not a large raw
persistence-over-peak gap: at the trajectory level the absolute AUCs are close (persistence 0.792 vs
peak 0.750 over the four promoted survivors at drift 0.4295), because grafts sitting near tau can lift peak as well. Here, peak-delta
does not separate attack from sham while persistence-delta does, which is the evidence for temporal
accumulation rather than one hot-ish graft per trajectory.

The result is explicitly descriptive, not powered. Separation is partial: two benign sham-deltas
($+0.0597$, $+0.0545$) exceed two attack survivor-deltas ($+0.0082$, $+0.0329$); the win rides on two
carriers ($+0.143$, $+0.051$). The cooled benign-sham rule both makes the bar honest and lowers drift,
which helps grafts clear it; the effect is real and disclosed.
We report these AUCs as descriptive point estimates rather than with confidence intervals: at four
attack survivors and six benign shams an interval would span nearly all of $[0,1]$, and the
persistence-delta-versus-peak-delta contrast is likewise not significance-tested; Section~\ref{sec:powering}
is the formal statement of why this axis cannot be powered.

\subsection{Result 2: Broadened Actions Generalization Failure and Sensor Margin}

Broadening the same density-four construction to clean file/download/path actions and a second
(shopping) scenario produces a decisive null (gate drift 0.4285):

\begin{itemize}
\item \textbf{persistence-delta AUC 0.167}, attack survivor median \textbf{$+0.0$} vs benign sham \textbf{$+0.0235$}.
\item all three survivor persistence-deltas exactly 0.0 (grafts added zero persistence mass).
\item 0\% of survivor grafts reach the productive 0.46--0.50 band (graft $s_i$ median 0.410, max 0.431).
\item half of the attack carriers (3 of 6) rejected by the directive floor.
\item a pre-registered native counterfactual was healthy before the run, so the null is not a selection
  artifact: native-only peak AUC was 0.653 and native-only persistence AUC was 0.556, with attack
  native persistence median 0.0354 versus benign 0.0434.
\item absolute post-graft AUC (three promoted survivors, drift 0.4285): peak 0.722 versus persistence
  0.556, peak exceeding persistence (the reverse of the git axis) because the grafts banked no
  persistence mass while sitting near enough to tau to lift peak. The post-graft persistence here
  (0.556, survivor roster) is distinct from the native-only persistence 0.556 above, which is the
  separate six-versus-six pre-graft roster; the digits coincide.
\end{itemize}

The transfer gate also stayed clean in the narrow sense. Of 23 attempted attack grafts, none leaked over tau under \Deval{}, and no trajectory was relabeled concentrated. One of the six carriers was authored with three grafts rather than four because of a generation shortfall, not an edge limit; that carrier offered 13 graftable action edges. The failure was the
opposite of a transfer overshoot: promoted attack grafts stayed sub-threshold but landed too cold
under \Deval{} to bank persistence.

A per-graft audit locates the failure precisely. On the broadened axis the grafts are elevated under
the \emph{generation target} (the \Dgen{} score, written $\shat$, sits uniformly near 0.42) but \textbf{collapse
onto the benign distribution under the detector of record}: attack $s_i$ $\sim$0.39--0.43 vs benign sham
$\sim$0.42--0.45, overlapping, with some benign
carrier actions (github reads, \texttt{get\_repository\_info}) scoring \emph{hotter} than the attack download
directives. \Deval{} assigns a near-constant $\sim$0.42 to almost every action on this axis, attack and
benign alike, so there is no per-event margin to bank.

A sweep of the CUSUM drift across its full range ($k = 0.25$ to $0.50$) finds \textbf{no operating point} at
which the attack grafts out-accumulate the benign shams: at every productive drift the benign shams
dominate. The lone apparent absolute-persistence separation (AUC 0.889 at drift 0.25) is a trap: not
graft-attributable, driven by native trajectory length and selection heat, at an operating point
that floods the benign population with persistence mass (it would detect ``long github task,'' not the
injection).

\textbf{The broadened-axis failure is therefore a sensor-margin failure, not a drift/CUSUM threshold
failure.} A CUSUM can amplify an existing per-event margin but cannot manufacture one. This bounds
the positive result to the git repo-exfil axis, where a genuine per-event margin exists, and shows
the productive band is axis-specific. (A supporting inversion reinforces the point: on this axis, the three floor-rejected attacks contributed more persistence than all three promoted survivors, each of which banked exactly 0.0. Here, the directive floor and persistence contribution are anti-correlated.)

\subsection{Result 3: Independent Powering Is Structurally Blocked}\label{sec:powering}

Powering the git-axis positive runs into a structural ceiling:

\begin{itemize}
\item The git repo-exfil axis exposes only \textbf{four exfil-shaped user tasks} (clone/push targets), of
  which \textbf{three} carry the full density-4 executed-action geometry the endpoint needs (the Result 1
  positive drew its four survivors from those three). Pairing different model executions of those
  same tasks raises the nominal sample (e.g.\ four models over three tasks $=$ N$=$12) but yields
  \textbf{pseudo-replicates}, not independent evidence: shared task structure, native heat, and directive
  geometry.
\item A pre-generation projection from the observed per-task deltas shows even an optimistic powered run
  reaches a task-clustered AUC of only $\sim$0.78, carried largely by a single task. The deeper problem is
  the test itself: the proposed task-level rank test has only three attack task means against six
  benign sham deltas. Under \emph{perfect} attack-versus-benign separation, the exact one-sided floor is
  $1 / \binom{9}{3} = 0.0119$; any imperfect separation is worse. The projected effect is not perfect, so
  a clustered N$=$3 cannot support a robust powered claim, even though the best-case perfect-separation
  floor sits nominally below 0.05.
\item A pre-registered native-counterfactual gate on the proposed powered roster then \emph{failed} (native
  persistence AUC 0.292 against the near-chance acceptance band 0.45--0.55, with matched benign
  controls running hotter than attack), and reaching the target sample forced reuse of prior
  carriers.
\item A read-only source scout confirmed the ceiling is external too: the second staged benchmark has no
  git suite, the train/eval split adds no new tasks, and no off-the-shelf source supplies the
  required typed git-exfil action edges with clean benign backbones and genuinely new tasks
  (real-git sources express git as freeform shell strings, not bindable typed edges).
\end{itemize}

Powering thus requires new data engineering or new task authoring, neither of which is a ready
experiment. We report Result 3 as a structural source-scarcity finding: adding more executions of
the same few tasks would increase nominal N without increasing independent evidence.

\subsection{The narrow productive band (why density is required, and why it is fragile)}\label{sec:band}

An earlier and larger run, predating the density-4 construction of Result 1, makes the mechanism's
arithmetic explicit even though it supplies no independent task-level power. In it we steered grafts
into a \emph{comfort band} deep below tau ($s_i$ 0.41 to 0.46). This succeeded at staying sub-threshold but
\textbf{flattened the accumulator}: with 9
attack and 6 benign trajectories at drift 0.451, persistence AUC 0.463 vs peak AUC 0.454, both at
chance. The cause is mechanical, not an instrument fault: the CUSUM banks only mass above the benign
drift ($\sim$0.451), so two grafts sitting \emph{at or below} drift in a 7--24 event trajectory contribute
$\sim$zero persistence. Contrast the earlier run whose grafts sat at 0.47--0.49 (above drift, hugging
tau) and drove persistence AUC 0.75.

This is the decisive design tension: \textbf{deep-sub-tau and out-persists-benign are in direct conflict
at low graft counts.} Two bands are at play here and they are not the same. The \emph{survivable-and-visible} band runs from the benign drift up to tau (roughly 0.45 to 0.50, given drift $\sim$0.43--0.45 and
tau 0.50): below drift the CUSUM banks nothing, at or above tau the event leaks, so this $\sim$0.05-wide
window is the mechanism's entire operating envelope. Within it we \emph{target} a tighter productive band,
0.46 to 0.50, deliberately offset above drift so grafts bank persistence mass rather than hugging the
drift floor; this is the band the per-carrier landing fractions in Results 1--2 (25\% vs 0\%) are
measured against. Result 1's positive comes
from \emph{density}: more mild grafts per trajectory let sustained elevation accumulate without any single
event hugging tau, but only when the detector places those grafts above drift in the first place,
which is exactly what fails on the broadened axis (Result 2). The productive band is the mechanism's
whole operating envelope, and it is narrow.

\subsection{Cross-family transfer accounting (both density runs)}

Anti-circularity is a transfer claim (Section~\ref{sec:method}), so we report transfer as a measured gate quantity
rather than an assurance. A graft is cloaked sub-threshold against \Dgen{} during construction; its
low-per-event-signal property is then required to \emph{survive re-scoring} under the different-family
\Deval{}, which the generator never sees. The table below makes that survival auditable for both density
runs: how many attack grafts were attempted, how many leaked (scored at or above tau under \Deval{}),
the leak fraction, how many trajectories were relabeled or floor-rejected as gate outputs, how many
attack survivors entered the slow-burn endpoint, and where the grafts landed under \Deval{}.

\begin{table}[h]\centering
\resizebox{\textwidth}{!}{%
\begin{tabular}{lrrrrrrrrr}
\toprule
run (axis) & gate drift & attempted & leaks ($s_i\!\ge\!\tau$) & leak frac &
  relabels & floor rej. & survivors & $s_i$ med. & $s_i$ max \\
\midrule
git repo-exfil (R1)   & 0.4295 & 24 & 1 & 0.042 & 1 & 1 & 4 (3 tasks) & 0.437 & 0.523 \\
broadened clean (R2)  & 0.4285 & 23 & 0 & 0.000 & 0 & 3 & 3           & 0.416 & 0.471 \\
\bottomrule
\end{tabular}%
}
\caption{Cross-family transfer accounting over all attempted cloaked-live attack grafts.}
\label{tab:transfer}
\end{table}

Medians and maxes are over all attempted cloaked-live attack grafts (the population the leak and
relabel counts also range over); the survivor-only graft medians quoted in Results 1--2 (0.434 for
the git-axis run, 0.410 for the broadened run) are the subset that entered the endpoint. Both runs
are \emph{non-catastrophic transfer}: in the git-axis run only one of 24 grafts crossed tau under the
cross-family evaluator (that single leak is the $s_i$ max 0.523, and it triggered the one concentrated
relabel), so the cloaking targeted at \Dgen{} mostly held under \Deval{} without ever being fitted to it.
The contrast with the broadened run is the story of Result 2: it leaked nothing, with 0 of 23 grafts crossing threshold, but its grafts landed too cold to matter. The maximum score was 0.471, below tau, and the median was 0.416, essentially at drift. Transfer succeeded only in the trivial sense of staying sub-threshold; it failed in the sense that mattered, leaving no per-event margin to accumulate. Transfer being non-catastrophic is necessary for the mechanism question to be
non-circular, but, as the broadened run shows, it is not sufficient for the accumulator to work.

\section{Methodological notes}

\subsection{The pseudo-replication trap in agent-benchmark evaluation}\label{sec:pseudorep}

The most reusable warning here is negative. Many agent-evaluation papers would run four models over
three tasks and report N$=$12. That sample is \textbf{not} twelve independent draws: the four executions of
one task share task structure, native trajectory heat, and directive geometry, so they are
correlated pseudo-replicates. Resampling at the carrier level overstates power; the bootstrap must
resample at the genuine independence unit (here, the \emph{task}, 3--4 units) and the paper must \emph{declare}
that unit. This distinction is what separates Result 0 (trajectory-unit, no pseudo-replication, a
legitimate n$=$9 bootstrap) from Results 1--3 (task-unit, where a carrier-level bootstrap would
manufacture significance). We treat this as a reusable reporting lesson: pseudo-replication is a
systematic way agent benchmarks overclaim power, and naming the independence unit should be a
reporting requirement.

\subsection{An independence ladder, and a worked algorithmic-independence audit}\label{sec:ladder}

A result computed under one detector version and reported against another invites a fair ``same
detector?'' challenge. We reconcile by re-deriving the headline numbers from a frozen per-event
receipt without re-running the model, and we declare \emph{which rung of independence} the verification
reached:

\begin{itemize}
\item \textbf{Rung 1, data-path independence (held).} The verifier re-derives every number from the raw
  per-event receipt, not the reported AUCs.
\item \textbf{Rung 2, algorithmic independence (validated).} Two code-disjoint \emph{blind} reimplementations of
  the accumulator/AUC math, written from a math-only spec alone (neither having read the canonical
  code, the verifier's re-derivation, or any expected value), reproduce all four AUCs and the drift
  constant to 6 decimal places (drift 0.437291, foil 0.766338, peak 0.827778, persistence 0.858333,
  frac 0.656944). An in-house re-derivation written \emph{after} reading the canonical code is only an
  anchored consistency guard (a reproduced latent bug would also match), and we say so.
\item \textbf{Residual, scoring-layer trust (open, bounded not closed).} The per-event scores are
  trust-on-faith from the model-half; the verifier never runs the model. The pins (BGE model id,
  engine git SHA, bank sha, fed-payload sha) bound this residual but do not close it. We name it.
\end{itemize}

A second, generalizable discipline emerges from the rung-2 audit: \textbf{a blind diff validates only the
code paths the data actually exercised.} A three-way match is only meaningful on a path the data
forces all three implementations to take; on a path the data never reaches, agreement is vacuous
because no implementation ran it. A clean overall match is therefore not blanket validation, and we
audited which of the spec's four candidate ambiguities the data genuinely tested.

Two were exercised by the data itself. The first is Mann-Whitney tie-handling: \texttt{frac\_auc} had 125 tied
pairs out of 360, and its rank-sum numerator came out to the half-integer 236.5. That half-integer is the proof: it can only arise if tied pairs were summed at half-weight. Three implementations landing on it identically therefore shows that they handled ties the same way, not that they merely agreed on an unstressed path. The second is CUSUM initialization $g_0 = 0$: 26 of 49
trajectories change persistence under an alternative init ($g_0 = -k$), so the reset-at-zero choice is
load-bearing rather than washed out, and agreement on it is again a forced match. The other two ambiguities, event ordering and even-N median drift, were dark on this dataset because the receipt was already position-sorted and the benign pool had an odd count. We therefore exercised them deliberately with differential fixtures that would diverge under an incorrect implementation.

All three code-disjoint implementations agreed across both exercised paths and both fixtures. This does
not prove the spec globally unambiguous; it shows that on every code path this study could exercise,
the spec resolved without reference to our codebase. We promote this rule to a charter-level reporting standard: enumerate which paths were exercised and which remained dark, then either fixture or carry each dark path.

We apply the same ladder to the Result 1 matched peak-delta control. It is emitted into the frozen native-delta receipt on the persistence endpoint's within-carrier, grafts-removed baseline, which is rung 1. An independent blind reimplementation, written from a math-only spec without sight of the canonical readout, its receipt, or any expected value, then reproduces the load-bearing quantity, peak-delta AUC, to full float precision, which is rung 2. The two
descriptive medians agree only under a declared canonicalization: the readout rounds each per-carrier
delta to 4 dp before the median, and the AUC is rank-based and so invariant to that rounding while the
medians are not. The Mann-Whitney engine carries the three-way validation above; the peak-delta's own
agreement is two-way (blind reimplementation versus the canonical readout), which we state rather than
round up. The readout and blind script are
run-local artifacts released with the public bundle (Appendix~\ref{sec:appendix}), not version-controlled with the
engine; reproducibility rests on the tracked accumulator plus the released score-only receipt.

\section{Discussion}

\textbf{Accumulation is not magic.} It is a margin amplifier, not a semantic detector. Where the per-event
detector produces elevated sub-threshold signal on the attack steps (git repo-exfil), accumulating
that signal across a trajectory recovers what per-event scoring misses. Where the detector assigns
attack and benign actions the same $\sim$0.42 (broadened clean-path actions), accumulation has nothing to
amplify, and no CUSUM drift recovers a margin that does not exist. The generation loop made this
concrete: it produced grafts that stayed elevated under the autoregressive cloaking target but failed
to transfer as elevated events under the held-out linear/retrieval evaluator. The failure is in the
sensor's per-event semantics, not in the temporal layer.

\textbf{The negative results are the contribution boundary, not a weakness.} ``Temporal accumulation
amplifies a weak margin; it cannot create one'' is a falsifiable mechanism claim that tells a
practitioner exactly when proxy-side temporal evidence will and will not help. Result 3 is also a
community warning about pseudo-replication: model reruns over the same task geometry should not be
counted as independent evidence.

\textbf{Why graft-delta, not absolute AUC.} Upgrading from the distance member to the contrastive member
raised the per-event foil ($0.677 \to 0.766$) and thereby narrowed the absolute peak-to-persistence gap.
A headline built on absolute persistence-over-peak AUC would have moved with the detector version.
The persistence-delta isolates the accumulator's added value while neutralizing native carrier heat,
and the peak-delta control checks whether the same grafts merely raised the hottest event. We
recommend paired selection-controlled deltas over absolute AUC for any accumulation claim of this
shape.

\section{Limitations}

\begin{itemize}
\item The positive rests on \textbf{three independent tasks and four survivors}. It is a mechanism
  demonstration, not a measured effect size.
\item The productive band that makes density pay off is \textbf{axis-specific} (git repo-exfil) and was not
  found elsewhere under this detector.
\item The native selection confound, although cancelled by the graft-delta endpoint, is \textbf{material in
  absolute terms} and task-locked.
\item All results are conditioned on \textbf{one frozen detector of record and one floor judge}. A different
  sensor could move the axis boundary; that is exactly why we did not tune it.
\item The raw persistence-over-peak AUC margin is \textbf{small even where density works} (Result 0: 0.858 vs
  0.828, statistically tied; Result 1: persistence 0.792 vs peak 0.750) and \textbf{reverses where it fails}
  (Result 2: peak 0.722 $>$ persistence 0.556; Results 1 and 2 on their promoted-survivor roster at the
  endpoint drift), because grafts sitting near tau lift peak too. The mechanism claim
  therefore rests on the controlled persistence-delta and its matched peak-delta control, not on
  persistence dominating peak in absolute terms. We do not claim a persistence-beats-peak headline.
\item The \textbf{scoring-layer trust boundary} (Section~\ref{sec:ladder}) is bounded by pins but not closed: a verifier
  re-derives the arithmetic, not the model scores.
\end{itemize}

\section{Future work}

The most important missing experiment is not another rerun over the same carriers, but a genuinely
new typed-action source. Reusing model executions over the same tasks would inflate nominal N without
adding independent evidence. The decisive next test is therefore either newly authored git-exfil
tasks with the same typed-edge semantics, or an extractor that converts real git-agent traces into
bindable typed action edges. That experiment would test whether the git-axis margin observed here is
a property of the temporal-accumulation mechanism or an artifact of this task pool.

A second direction is sensor variation. The broadened-axis null appears because \Deval{} assigns
attack and benign clean-path actions nearly the same per-event scores; a different detector or floor
judge could move that boundary. We treat that as a separate program rather than a hidden degree of
freedom in this paper: changing the sensor would answer a different question, namely which per-event
semantics make temporal accumulation useful.

Finally, the accumulator family itself is intentionally simple. Peak and reset-at-zero CUSUM make the
mechanism auditable, but other temporal statistics could trade sensitivity against false positives in
different ways. Those variants should be evaluated only after a larger independent task source exists;
otherwise accumulator tuning would be dominated by the same source-scarcity and pseudo-replication
issues identified here.

\section{Related work}

\textbf{Prompt injection and agent security.} Direct prompt injection, adversarial instructions that override a model's intended task, was characterized early by Perez and Ribeiro in ``Ignore Previous Prompt''~\citep{perez2022promptinject}. Greshake et al., in ``Not What You've Signed Up For''~\citep{greshake2023indirect}, later showed the \emph{indirect}
variant, where the payload arrives through retrieved or tool-returned content, is a central risk
for tool-using agents. Liu et al.\ (USENIX Security 2024~\citep{liu2024formalizing}) formalize and benchmark injection attacks
and defenses across tasks and models. Li et al.\ introduce Multi-Turn Human Jailbreaks (MHJ)~\citep{li2024mhj}, which
we use as three held-out control-plane conversations in the foundation attack cohort.
Agent-security benchmarks supply the trajectories such defenses are measured on: AgentDojo
(Debenedetti et al., NeurIPS D\&B 2024~\citep{debenedetti2024agentdojo}) and InjecAgent (Zhan et al., 2024~\citep{zhan2024injecagent}) instrument tool-using
agents with attacker-controlled content, and AgentDyn (arXiv 2602.03117~\citep{agentdyn2026}), on which our carriers
build, exposes the typed action edges we graft onto. Many defenses score a single event, message, or
prompt context -- prompt-injection classifiers, known-answer probes, and prompt-level mitigations
such as spotlighting (Hines et al., 2024~\citep{hines2024spotlighting}). Our contribution is orthogonal to this line: we hold the
per-event detector frozen and ask whether a \emph{trajectory-level} accumulator recovers signal the
per-event view structurally misses, rather than proposing a stronger per-event detector.

\textbf{Sequential change detection.} The accumulator is a CUSUM (Page, 1954~\citep{page1954cusum}), the classical
change-point statistic, with relatives in Shewhart~\citep{shewhart1931economic} and EWMA~\citep{roberts1959ewma} control charts and in the
quickest-change-detection tradition (Lorden, 1971~\citep{lorden1971procedures}; Pollak, 1985~\citep{pollak1985optimal};
treatments by Lai, 1998~\citep{lai1998information} and Poor and Hadjiliadis, 2009~\citep{poor2009quickest}). Our use is non-standard in two ways: the ``change'' is a slow, distributed
elevation rather than a single regime shift, and the drift constant is \emph{benign-referenced} (the
median per-event score over the benign population) so the statistic banks only mass a real benign
trajectory would not. The result is a proxy-side trajectory monitor, not an online detector with a
calibrated false-alarm rate; our Section~\ref{sec:band} arithmetic for the productive band is the CUSUM
counterpart of an ARL/threshold analysis, specialized to the slow-burn regime.

\textbf{Non-circular evaluation.} Our anti-circularity discipline draws on two literatures. From
adversarial robustness, a repeated lesson is that evaluations collapse when the attack can see or fit the defense. This lesson appears in obfuscated gradients~\citep{athalye2018obfuscated}, adaptive attacks~\citep{tramer2020adaptive}, and broader robustness-evaluation guidance~\citep{carlini2019evaluating}. It motivates our held-out generator/evaluator split and the rule that the evaluated property must transfer by re-scoring, never by optimization against the evaluator. The transferability of adversarial examples across model
families (Papernot et al., 2016~\citep{papernot2017blackbox}; Demontis et al., USENIX Security 2019~\citep{demontis2019transfer}) is the closest adversarial-ML
analogue to the property we test and machine-check: \Dgen{} and \Deval{} are disjoint families, so
survival under \Deval{} is transfer, not fitting. From the LLM-evaluation side, data-contamination and leakage concerns motivate
our reconciliation step; we contribute an explicit \emph{independence ladder} (data-path, algorithmic,
scoring-layer) and a worked blind-reimplementation audit as a reporting standard, distinguishing code
paths the data exercised from those it left dark.

\textbf{Pseudo-replication.} The independence-unit problem is the statistical-ecology notion of
pseudo-replication (Hurlbert, ``Pseudoreplication and the Design of Ecological Field Experiments'', 1984~\citep{hurlbert1984pseudoreplication}),
revisited for the life sciences by Lazic (2010~\citep{lazic2010pseudoreplication}). The same error recurs in multi-model
agent-benchmark evaluation: running several model executions over a handful of tasks and resampling at
the execution level treats correlated replicates as independent draws. We import the discipline of
naming and resampling at the genuine independence unit (here, the task) and argue it should be a
reporting requirement for agent-benchmark power claims.

\section{Conclusion}

A proxy-side temporal accumulator can recover distributed, individually sub-threshold control-plane
injection signal that per-event detection misses, but only where the underlying per-event sensor
already has a real margin on the attack steps. Under a fully frozen, non-circular, cross-family test,
the mechanism is real on git repo-exfil (raw attack persistence-deltas $+0.0513$, $+0.0329$, $+0.1432$,
$+0.0082$; persistence-delta AUC 0.708 over four attack survivors and six benign shams; matched
peak-delta AUC 0.417),
fails to generalize to broader clean-path actions because the sensor margin collapses there
(persistence-delta AUC 0.167, a sensor-margin failure not a threshold failure), and cannot be
independently powered because the axis exposes only three to four independent tasks. We frame
accumulation as a margin amplifier, give the narrow productive-band condition under which density
pays off, and include a pseudo-replication warning plus a worked algorithmic-independence audit for
agent-benchmark evaluation. This is a bounded mechanism finding, reported as such, not a deployable
detector.

\appendix
\section{Reproducibility and provenance}\label{sec:appendix}

\textbf{Frozen stack.} \Dgen{} $=$ served \texttt{Qwen3-30B (MLX 4-bit, v2507)}, cache-off (deterministic). \Deval{} of
record $=$ ensemble-max over \{ L1 generalist (linear), contrastive embedding member over
\texttt{contrastive reference bank v3}, sha \texttt{15302fcde748} \}. Encoder $=$ \texttt{BAAI/bge-small-en-v1.5} (MIT, CLS
pooling, 384-d). Accumulator $=$ peak $+$ persistence\_cusum. \tauev{} 0.50, $\epsilon$ 0.05, \texttt{max\_iter} 6
(\texttt{calibration/tau\_event\_lock.json}). Floor $=$ held-out \texttt{GPT-5-Codex}, per-graft, fail-closed. Engine
git SHA at scoring \texttt{abf36d4}. The contrastive member uses its own committed sigmoid ($a=23.72$,
$b=0.16$), not the retired distance member's tau-lock sigmoid.

\textbf{Operating-point lineage.} The original 2026-06-07 foundation used the retired distance member
(reference sha \texttt{164a9e12}, sigmoid $a=6.326$/$b=0.7421$; foil 0.677 / peak 0.789 / persistence 0.814).
The whole paper was reconciled onto the single contrastive detector of record (foil 0.766 / peak
0.828 / persistence 0.858), verified across the compute boundary; verdict PASS. The pass criterion is
committed in tracked code at \texttt{fc8a45e} (2026-06-07), three days before the contrastive member existed
(\texttt{f63e5da}, 2026-06-10), so it could not have been reverse-engineered from the contrastive gap.

\textbf{Benchmark.} AgentDyn \citep{agentdyn2026} (MIT, built on AgentDojo). Foundation attack cohort: 6
handcrafted multi-turn attacks $+$ 3 MHJ control-plane conversations, held out of the reference banks.

\textbf{Reproducibility boundary.} The public release should include the detector/accumulator code, the
frozen configuration, the non-data-bearing receipt schema, the blind reimplementation scripts, and
synthetic dark-path fixtures for tie handling, event ordering, CUSUM initialization, and even-N
median drift. The per-event receipts that contain trajectory text or graft payloads are not part of
the public artifact because they may contain private or benchmark-restricted content. The headline
AUCs and bootstrap gaps are nevertheless model-free once the per-event score receipt is fixed: a
verifier can recompute them from rows of \texttt{(trajectory\_id, group, position, s\_i)} plus the declared
drift rule, without access to the model server or generator.

\textbf{Audit trail.} Internally, every reported number is backed by immutable run receipts and an
independent model-free reconciliation pass. For the public paper, we report the hashes, frozen stack,
score schema, and reconciliation procedure rather than internal notebook or run-directory paths. If
the underlying benchmark license permits, a redacted score-only receipt can be released separately;
otherwise, the reproducible public artifact is the arithmetic verifier plus fixtures, and the
scoring-layer receipt remains an explicitly named trust boundary.

\textbf{Roster provenance.} Two distinct carrier rosters back the reported AUCs and must not be conflated
when reproducing. The native-counterfactual AUCs (git axis: native peak 0.778, persistence 0.750) are
computed over the pre-graft attack roster (six attack versus six benign carriers); the graft-delta
endpoints are computed over the post-gate survivor set (four attack survivors versus six benign shams).
Recomputing a native-counterfactual AUC from the survivor-set receipt returns a different value (peak
0.750 rather than 0.778) because the roster differs, not the arithmetic. Each native-counterfactual
number resolves against its own carrier-selection receipt; each delta endpoint against its survivor-set
receipt.

\textbf{Ethics, authorization, and responsible use.} All attacks in this work were constructed on a public
agent-security benchmark (AgentDojo / AgentDyn) and were never deployed against any production system or
third party. Adversarial payload generation (the cloaking target \Dgen{}) used a locally served
open-weight model (Qwen3-30B), so no third-party hosted provider was asked to produce attack content.
The only hosted commercial model in the evaluated pipeline is the directive-floor judge (GPT-5-Codex),
used for verification rather than generation; the detector of record (\Deval{}) is a frozen local linear
SVM plus open contrastive encoder, so no commercial model performs detection. Commercial-model use
complied with the provider's usage policy and the authors' participation in OpenAI's Trusted Access for
Cyber program.

\bibliographystyle{plainnat}
\bibliography{refs}

\end{document}

%% file: local_author.tex
\author{J Alex Corll}